\documentclass[sn-mathphys-num]{sn-jnl}% Math and Physical Sciences Numbered Reference Style 
\usepackage{subfig}
\usepackage{url}
\usepackage{xcolor}
\usepackage{array}
\usepackage{verbatim}
\usepackage{graphicx}
\usepackage{graphicx}%
\usepackage{multirow}%
\usepackage{amsmath,amssymb,amsfonts}%
\usepackage{amsthm}%
\usepackage{mathrsfs}%
\usepackage[title]{appendix}%
\usepackage{graphicx}
\usepackage{subfig}
\usepackage{xcolor}%
\usepackage{textcomp}%
\usepackage{manyfoot}%
\usepackage{booktabs}%
\usepackage{algorithm}%
\usepackage{algorithmicx}%
\usepackage{algpseudocode}%
\usepackage{listings}%
\usepackage{algpseudocode}
\usepackage{caption}

\theoremstyle{thmstyleone}%
\theoremstyle{thmstyletwo}%

\theoremstyle{thmstylethree}%

\begin{document}

\title[Article Title]{Fault Detection in Electrical Distribution System using Autoencoders}


\author[1,2]{\fnm{Sidharthenee} \sur{Nayak}}
\author[1,3]{\fnm{Victor} \sur{Sam Moses Babu}}
\author[2]{\fnm{Chandrashekhar} \sur{Narayan Bhende}}
\author[3]{\fnm{Pratyush} \sur{Chakraborty}}

\author*[1,]{\fnm{Mayukha} \sur{Pal}}\email{mayukha.pal@in.abb.com}

\affil[1]{\orgdiv{ABB Ability Innovation Center}, \orgname{Asea Brown Boveri Company}, \city{Hyderabad}, \postcode{500084}, \country{India}}

\affil[2]{\orgdiv{School of Electrical Sciences}, \orgname{Indian Institute of Technology Bhubaneswar}, \city{Bhubaneswar}, \postcode{751013}, \country{India}}

\affil[3]{\orgdiv{ Department of Electrical and Electronics Engineering}, \orgname{BITS
Pilani Hyderabad Campus}, \city{Hyderabad}, \postcode{500078}, \country{India}}


\abstract{In recent times, there has been considerable interest in fault detection within electrical power systems, garnering attention from both academic researchers and industry professionals. Despite the development of numerous fault detection methods and their adaptations over the past decade, their practical application remains highly challenging. Given the probabilistic nature of fault occurrences and parameters, certain decision-making tasks could be approached from a probabilistic standpoint. Protective systems are tasked with the detection, classification, and localization of faulty voltage and current line magnitudes, culminating in the activation of circuit breakers to isolate the faulty line. An essential aspect of designing effective fault detection systems lies in obtaining reliable data for training and testing, which is often scarce. Leveraging deep learning techniques, particularly the powerful capabilities of pattern classifiers in learning, generalizing, and parallel processing, offers promising avenues for intelligent fault detection. To address this, our paper proposes an anomaly-based approach for fault detection in electrical power systems, employing deep autoencoders. Additionally, we utilize Convolutional Autoencoders (CAE) for dimensionality reduction, which, due to its fewer parameters, requires less training time compared to conventional autoencoders. The proposed method demonstrates superior performance and accuracy compared to alternative detection approaches by achieving an accuracy of 97.62\% and 99.92\% on simulated and publicly available datasets. }




\maketitle

\section{Introduction}\label{sec1}

The electric power grid plays a vital role in modern society, reliably supplying electricity to residential, commercial, and industrial sectors. As our dependence on electricity grows, there is a corresponding increase in the need for robust and effective electrical distribution systems (EDSs). Guaranteeing the safety and dependability of these systems requires mitigating risks and ensuring uninterrupted power delivery.\cite{reddy2022data, DWIVEDI2023156} Therefore, the adoption of sophisticated methods for detecting and classifying faults is crucial to optimize the performance of EDSs.\cite{dwivedi2023evaluating, revEE} These techniques serve as a critical tool in identifying and managing faults, optimizing maintenance efforts, and ultimately strengthening the overall resilience of the grid \cite{dwivedi2023evaluating, P2P}. The electrical power system, comprising various dynamic elements, is susceptible to disturbances and faults, necessitating swift fault detection and protection operation to maintain stability \cite{jamil2015fault}. It is important to swiftly detect and classify faults on transmission lines, with protection systems initiating relays to prevent outages \cite{jamil2015fault}. Effective fault detection and classification, ensuring rapid restoration of the power system, are imperative for service reliability and minimizing outages \cite{singh2011transmission}. Protection schemes must promptly detect and remove affected segments during a fault incident to minimize its impact \cite{tirnovan2019advanced}. However, the expansion of modern power networks poses challenges for protection systems, requiring integrated schemes capable of monitoring different grid layers. Wide Area Protection (WAP) using phasor measurements from Phasor Measurement Units (PMUs) has been proposed, yet challenges remain in interpreting data and identifying faulty components \cite{ebrahim2019integrated}.

Existing fault detection algorithms for transmission networks often rely on iterative solutions or require numerous PMUs, while distribution networks face issues due to distributed generation impacting fault levels and relay operation \cite{mayukha2023data, dwivedi2023dynamopmu}. Synchrophasor measurements offer a more reliable alternative but are currently limited to distribution networks, highlighting the need for an integrated scheme applicable to both distribution and transmission networks \cite{ebrahim2019integrated}. 

Fault diagnosis is categorized into two main types: model-based and process history-based approaches. Model-based methods involve analyzing faults by representing a system or process using either quantitative or qualitative models. On the other hand, process history-based techniques rely on empirical data gathered from the process, establishing connections between inputs and desired outputs without prior mathematical modeling. Feature extraction is crucial in process history-based methods as it helps capture essential information from empirical data for pattern recognition. With advancements in signal processing and a deeper understanding of power systems, various techniques have emerged for direct measurement and transformation, enabling the extraction of inherent fault characteristics. Commonly utilized methods for feature extraction in the literature include Wavelet and Fourier transforms, which effectively isolate fault-related characteristics with robustness and precision. \cite{aleem2015methodologies,gu2003bridge,das2006comparison, abdollahi2010comparison}. However, these classical methods may yield inaccurate results due to assumptions about line
parameters. 
Artificial neural networks (ANNs) and support vector machines (SVMs) are robust pattern recognition methodologies capable of efficiently generalizing dynamic parameters using both supervised and unsupervised learning approaches. \cite{li2000application}. Recently, machine learning algorithms have been widely used by combining signal processing approaches to rapidly and accurately identify faults \cite{w1,ml3,ml5}. Signal processing techniques extract the features from initially captured voltage and current signals to determine fault occurrence and their types \cite{5}. Unfortunately, the selection of faulted features across different
frequency ranges is often arbitrary, leading to inconsistency in
results. Therefore, enhancing the fault detection accuracy for EDSs has
emerged as a significant research focus. Also, existing fault detection
techniques are supervised approaches, which poses challenges for
real-time applications due to the requirement of prior labeling,
and achieving online fault detection and clustering with a high
degree of accuracy remains elusive. Recently autoencoders have emerged as an interesting option for anomaly detection in a time series because it needs to be trained only on one type of data that is normal data. In \cite{chen2018autoencoder} authors have used deep autoencoders for anomaly detection in wireless communication networks. Similarly \cite{ribeiro2018study} have used autoencoders for anomaly detection in videos.

In our proposed method we have used a deep convolutional autoencoder model to detect faults in the distribution and transmission system.  At first, the model is trained on normal time-series data of current which has no fault. During training the autoencoder learns to reconstruct the normal time series data and the maximum reconstruction error is chosen as the threshold. While testing, the current signals with various types of faults are given to the model. If the reconstruction error is more than the threshold then those segments of the signal were identified as fault segments. The model was trained and tested on both simulated and publicly available datasets and gave high accuracy for fault detection in both datasets.

The key contributions of this paper are as follows:
\begin{enumerate}
    \item The work proposes the use of convolutional autoencoders for detecting faults in power systems.
    \item The autoencoder model performed better than other traditional ML models achieving a high accuracy for detecting faults. It achieved a notable accuracy of 97.62\% on the simulated dataset and 99.92\% on the publicly available dataset.
\end{enumerate}

The paper is structured as follows: section \ref{section:Methods} provides a detailed exploration of the autoencoder structure and how it is used for anomaly detection. Section \ref{section:Simulation} outlines information about the datasets used and evaluation metrics with a discussion of obtained results. Finally, section \ref{section:Conclusion} summarizes the paper.

\section{Materials and Methods}
\label{section:Methods}

The architecture of the proposed model for signal classification is illustrated in Fig. \ref{fig:en}.
% The model is first trained on data that does not contain any anomaly or fault. Then the trained model is used to reconstruct the training data and the maximum amount of reconstruction error on the training data is taken to be the threshold. Then the test signal which contains different kinds of faults is given to the model for reconstruction. The segments of the test signal for which the reconstruction error is more than the threshold are identified as fault segments. 

\begin{figure}[t]
\centering
\includegraphics[width=1.0\textwidth]{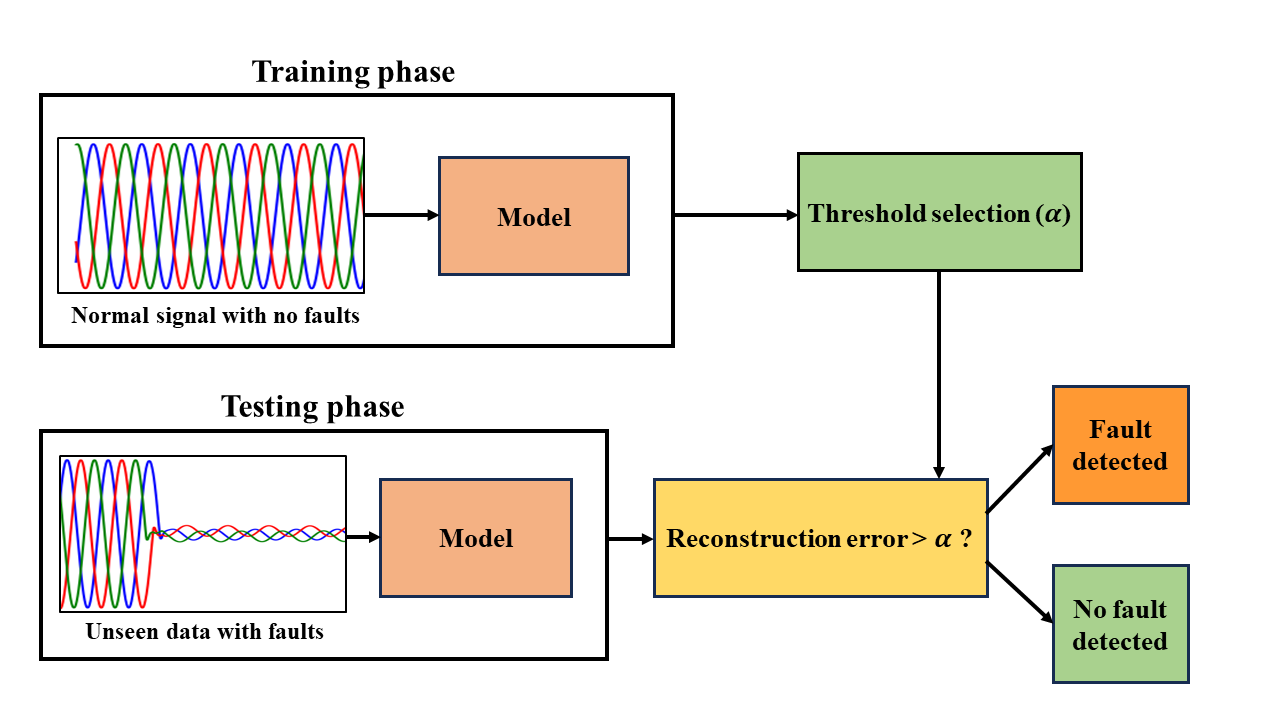}
  \caption{Flow chart of the proposed fault detection process}
  \label{fig:en}
\end{figure}

\begin{figure}[t]
\centering
\includegraphics[width=0.8\textwidth]{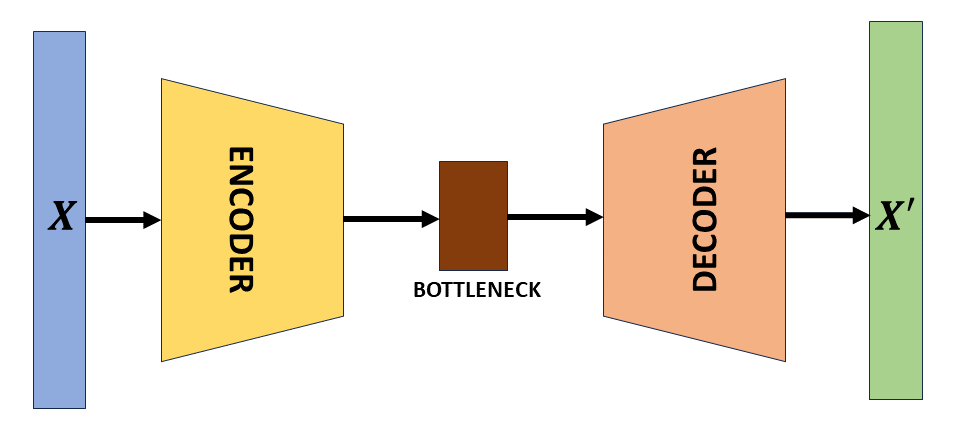}
  \caption{Architecture of autoencoder used in the proposed algorithm}
  \label{fig:en}
\end{figure}

\subsection{Autoencoders}
The architecture of an autoencoder is illustrated in Fig. \ref{fig:en}.
An autoencoder represents a specialized neural network designed primarily to compress input data into a meaningful representation and subsequently reconstruct it to closely resemble the original input.\cite{bank2023autoencoders}

Autoencoders were initially introduced by Rumelhart et al. \cite{rumelhart1985learning} as neural networks trained to reconstruct their input. They are primarily utilized for unsupervised learning to derive a meaningful representation of the data, which could be applied to tasks like clustering. Typically, the encoder is be formulated as a function dependent on certain parameters.
 
\begin{equation}
\mathbf{g}_i=A\left(\mathbf{x}_i\right)
\end{equation}

Where $\mathbf{g}_i \in \mathbb{R}^p$ is the output of the encoder block in Figure 1 when we evaluate it on the input $\mathbf{x}_i$.  The decoder (and the output of the network that we will indicate with $\tilde{\mathbf{x}}_i$ ) is be written as a second generic function of the latent features \cite{michelucci2022introduction}
\begin{equation}
\tilde{\mathbf{x}}_i=B\left(\mathbf{g}_i\right)=B\left(A\left(\mathbf{x}_i\right)\right)
\end{equation}

Where $\tilde{\mathbf{x}}_i \in \mathbb{R}^n$.

The problem, as formally defined in [2], is to learn the functions $A: \mathbb{R}^n \rightarrow \mathbb{R}^p$ (encoder) and $B: \mathbb{R}^p \rightarrow \mathbb{R}^n$ (decoder) that satisfy
\begin{equation}
  \arg \min _{A, B} E[\Delta(\mathbf{x}, B \circ A(\mathbf{x})]  
\end{equation}

Here, $E$ represents the expectation across the distribution of $x$, and $\Delta$ stands for the reconstruction loss function, quantifying the disparity between the decoder's output and the input. Typically, the input discrepancy is evaluated using the $\ell_2$-norm. In the most popular form of autoencoders, $A$ and $B$ are neural networks \cite{ranzato2007unsupervised}. 
 When $A$ and $B$ are linear operations, a linear autoencoder is obtained \cite{baldi1989neural}. In the scenario where non-linear operations are omitted, the linear autoencoder converges to the same latent representation as Principal Component Analysis (PCA) \cite{plaut2018principal}. Consequently, an autoencoder serves as an extension of PCA, as it will learn a non-linear manifold instead of merely identifying a low-dimensional hyperplane where the data resides.

The encoder reduces and extracts relevant features from the input and reduces the input dimension. The \emph{bottleneck} is the compressed representation obtained at the encoder output. The decoder decompresses the \emph{bottleneck} to give an output whose dimension is equal to the dimensions of the input. The loss function used in the training would be the error between the input given to the autoencoder model and the reconstructed output.

Autoencoders have the following features \cite{examplewebsite}:
\begin{itemize}
    \item They are tailored to specific datasets, implying their effectiveness in compressing data resembling what they were trained on. 
    \item They operate with loss, leading to a degradation in the quality of decompressed outputs compared to the original inputs, contrasting with lossless arithmetic compression.
    \item They possess the advantage of automatic learning from data samples, facilitating the training of customized versions optimized for particular input types without necessitating additional engineering efforts, solely relying on appropriate training data.
\end{itemize}

\subsection{Autoencoders for anomaly detection}    

The fundamental concept of autoencoder-based anomaly detection revolves around prioritizing the understanding of normal patterns rather than explicitly modeling anomalies. During training, the autoencoder is trained to accurately reconstruct data containing normal patterns, such as normal time series data, by minimizing a loss function assessing the fidelity of reconstructions. Subsequently, upon completion of training, the model excels in reconstructing data featuring normal patterns but struggles with anomalous data, as it hasn't been exposed to them during training. Anomaly detection is accomplished by evaluating the reconstruction metrics, like reconstruction error, which serve as indicators of anomalies. In this work, the autoencoder is trained on the current signal of a single phase under normal operating conditions. The reconstruction error between the predicted and original current signals is selected as the error threshold. Now for each phase, the current signals with faults are given to the model for prediction. The segments of the signal for which the reconstruction error is more than the error threshold are detected as an anomaly segment and hence the fault segment is detected.

\subsection{Model architecture}
\begin{figure*}
\centering
\includegraphics[width=1.0\textwidth]{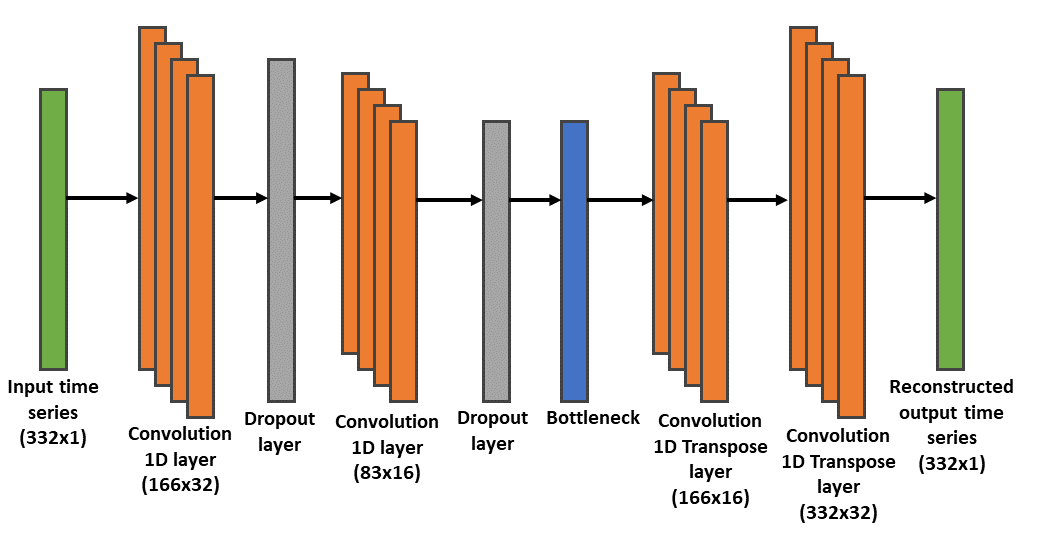}
  \caption{Detailed layers of the CNN-based autoencoder model}
  \label{fig:det}
\end{figure*}

In this paper we have used a Convolutional Neural Network (CNN) based architecture i.e., both encoder and decoder are CNN models. The entire input signal is divided into samples of fixed length using overlapping windows. Suppose the entire signal has $N$ data points, and if the fixed length is taken as $T$ then the entire signal is divided into segments of length $T$. So a total of $N-T+1$ samples are generated. Each sample is passed through the CNN-based encoder which creates a lower-dimensional representation of the input. This compressed representation is passed into the decoder which decompresses and regenerates. The detailed architecture of the encoder and decoder model is shown in Fig. \ref{fig:det}.

\subsection{Evaluation Metrics}
After the model detected the fault segments in the entire signal, it would classify each data point as faulty or not faulty. The overall performance of our model is evaluated from the confusion
matrix. There are four terms namely True
Positives (TP), False Positives (FP), False Negatives
(FN), and True Negatives (TN) are included in
this evaluation matrix. Evaluation indicators namely
accuracy, specificity, Recall, F1-score, and precision were evaluated to assess the performance of our model. \cite{deo2024supremacy}
These evaluation indicators are calculated
utilizing the formulas given in Table \ref{tab:eval_met}.

\begin{table}[]
\centering
{\renewcommand{\arraystretch}{1.5}
\caption{Evaluation metrics to assess model performance}
\begin{tabular}{lc}
\toprule
Metric  & Formula\\
\midrule
Accuracy \vspace{0.1cm} & $
        A_s = \dfrac{TP + TN}{TP + TN + FP + FN}$ \vspace{0.1cm}\\
Precision  \vspace{0.1cm} & $
        P_s = \dfrac{TP}{TP+FP}$   \vspace{0.1cm} \\
Recall \vspace{0.1cm} &  $
        R_s = \dfrac{TP}{TP+FN}$ \vspace{0.1cm}\\
Specificity \vspace{0.1cm} & $
        SF_s = \dfrac{TN}{TN+FP}$ \vspace{0.1cm}\\
F1-score \vspace{0.1cm} &    $
        F1_s = \dfrac{2 \times P_s \times R_s}{P_s+R_s}$ \vspace{0.1cm}\\
\bottomrule
\end{tabular} 
\vspace{2mm}
\label{tab:eval_met}}
\end{table}

\section{Dataset}
\label{section:Simulation}

\subsection{Simulation Model}
\begin{figure*}
\centering
\includegraphics[width=1.0\textwidth]{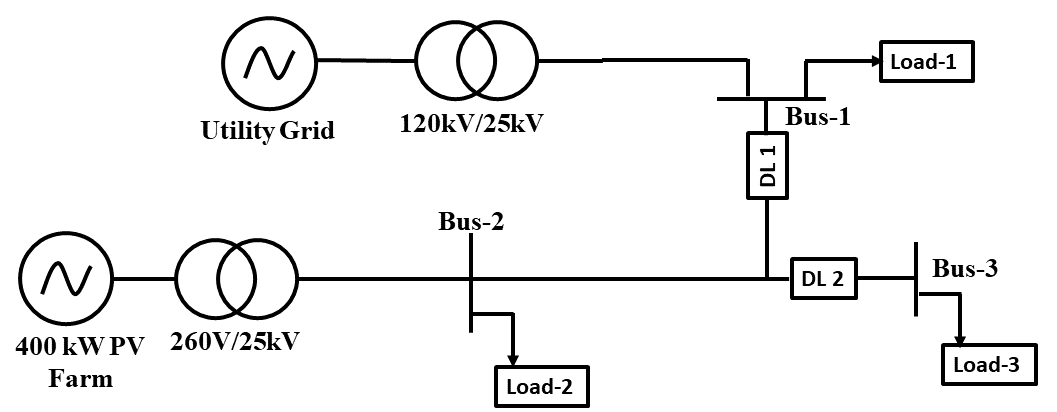}
  \caption{Single Line Diagram of simulation model consisting of a 400 kW solar PV farm connected to the grid with three loads and two feeders.}
  \label{fig:simulation}
\end{figure*}

\begin{figure*}[h!]
\centering
\captionsetup[subfigure]{font=small}
\subfloat[Current signals with no fault]{\includegraphics[width=0.48\textwidth, height=0.25\textwidth]{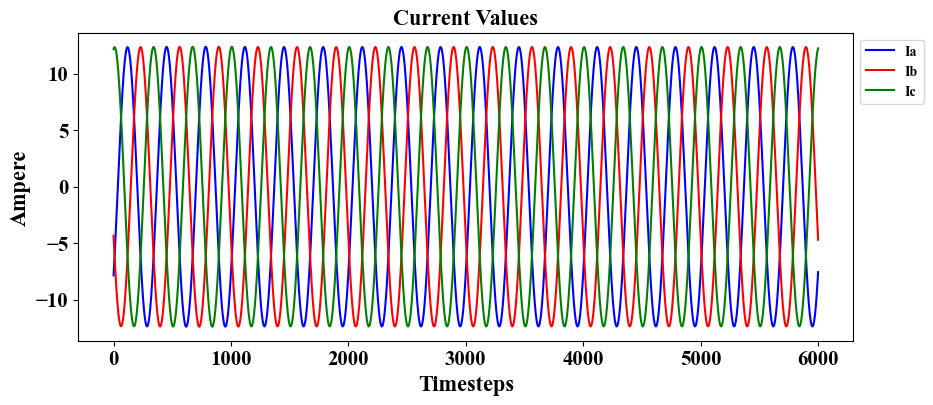} \label{fig:sim_sub1}}\;
\subfloat[Voltage signals with no fault]{\includegraphics[width=0.48\textwidth, height=0.25\textwidth]{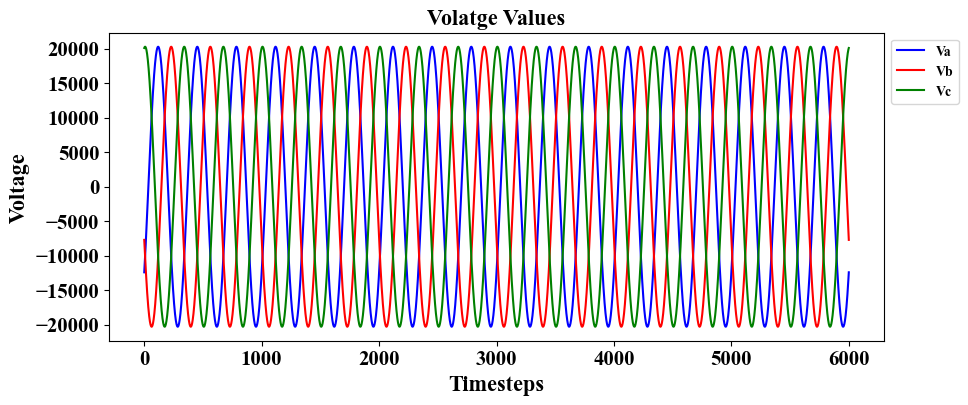} \label{fig:sim_sub2}}\;
\hfill
\subfloat[Current signals with four types of faults]{\includegraphics[width=0.48\textwidth, height=0.25\textwidth]{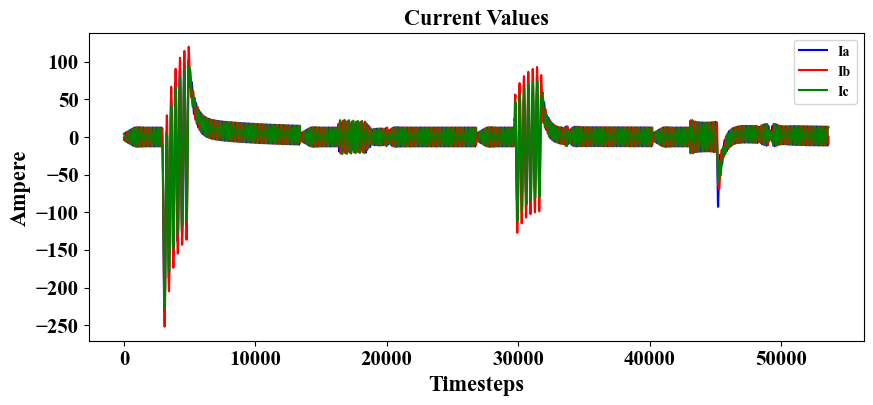}\label{fig:sim_sub3}}\;
\subfloat[Voltage signals with four types of faults]{\includegraphics[width=0.48\textwidth, height=0.25\textwidth]{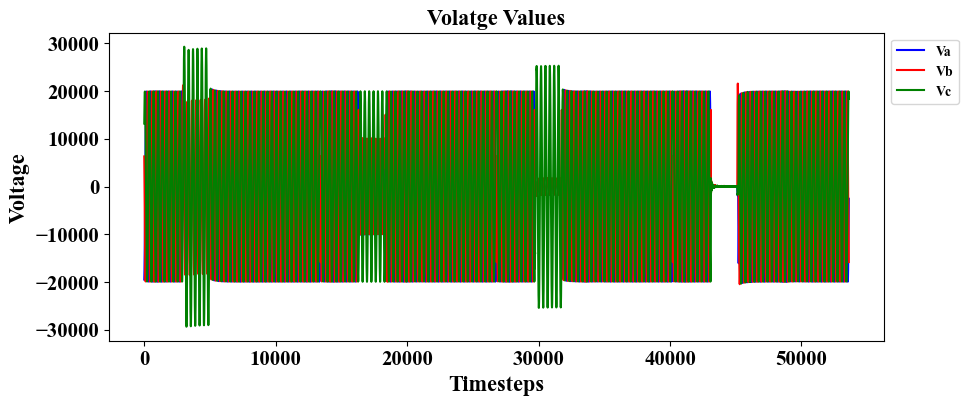}
\label{fig:sim_sub4}}\;

\caption{Current and Voltage signals of all three phases from the simulation model}
\label{fig:sim}
\end{figure*}

The schematic of the simulation model is shown in Fig. \ref{fig:simulation}. The simulation is performed in MATLAB/SIMULINK. The considered system consists of a solar PV farm of four 100 kW PV arrays connected to a distribution system. The distribution system consists of 2 feeders of 8 km and 14 km supplying 3 loads. There are 3 buses in the system and the PV farm is connected to Bus 2 where the three-phase voltage and current measurements are measured and given as input to the proposed model for fault detection. Four types of faults: single Line-to-Ground (LG) i.e., AG, double Line-to-Ground (LLG) i.e., ABG, triple Line-to-Ground (TLG) i.e., ABCG, and Line-to-Line i.e., AB faults were generated at a distance of 1 km with a fault resistance of 0.01 ohm. The sampling rate of the simulation is 5e\textsuperscript{-5} seconds i.e., 320 samples per cycle of 60Hz frequency. Each fault is simulated for 100 milliseconds i.e., 2000 samples. Using the three-phase voltage and current measurements at Bus 2, we perform the proposed methods for fault detection. Fig. \ref{fig:sim} shows the current and voltage signals of all three phases from the simulation model. Fig. \ref{fig:sim_sub1} and \ref{fig:sim_sub2}
shows the current and voltage signals with no fault. Fig. \ref{fig:sim_sub3} and \ref{fig:sim_sub4}
shows the current and voltage signals with four types of faults i.e., LG, LLG, TLG and, LL faults. 

\subsection{Public dataset}
We have also tested the model on a public dataset obtained from Kaggle \cite{data2}. The dataset originated from a power system simulated in MATLAB for fault-detection purposes. It is used for statistical comparision of the performance of our model with other traditional ML models.
% This power system comprises four generators operating at a voltage level of 11 kV, with each pair situated at opposite ends of the transmission line. Additionally, transformers are incorporated at intermediate points to replicate and investigate different faults occurring at the midpoint of the distribution line. 

\begin{figure*}[!h]
\centering
\captionsetup[subfigure]{font=small}
\subfloat[Current signals with different types of faults]{%
  \includegraphics[width=0.48\textwidth, height=0.25\textwidth]{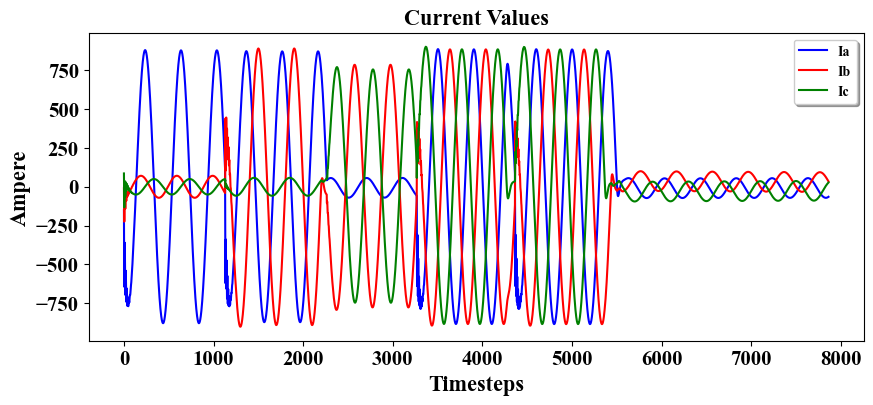}%
  \label{fig:sim_sub3}%
}
\subfloat[Voltage signals with different types of faults]{%
  \includegraphics[width=0.48\textwidth, height=0.25\textwidth]{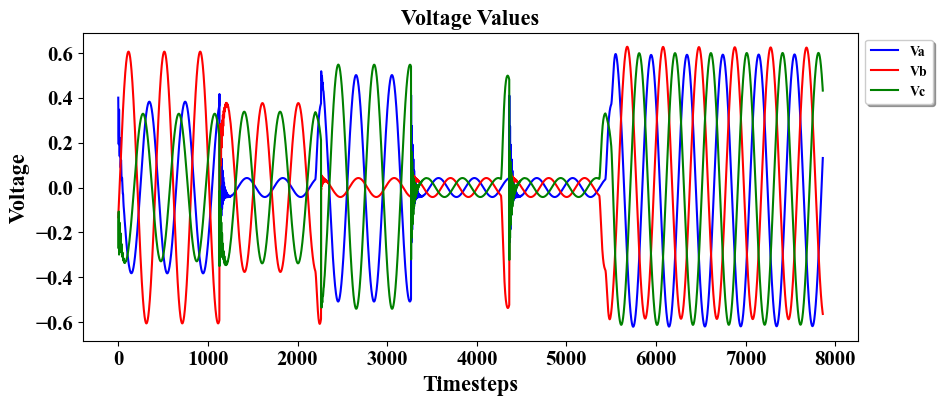}%
  \label{fig:sim_sub4}%
}
\caption{Current and Voltage signals of all three phases from the public dataset}
\label{fig:sim}
\end{figure*}

% \begin{figure}
%      \centering
%      \begin{subfigure}
%          \centering
%     \includegraphics[width=1.0\textwidth]{Figures/current_public.png}
%          \caption{Current signals with different types of faults}
%          \label{fig:sim_sub3}
%      \end{subfigure}
%      \hfill
%      \begin{subfigure}
%          \centering
%          \includegraphics[width=1.0\textwidth]{Figures/voltage_public.png}
%          \caption{Voltage signals with different types of faults}
%          \label{fig:sim_sub4}
%      \end{subfigure}
%         \caption{Current and Voltage signals of all three phases from the public dataset}
%         \label{fig:sim}
% \end{figure}

\section{Results and Discussion}

% The calculations were performed utilizing Python version 3.7.6, TensorFlow version 2.7.0, and Keras version 2.7.0 on a typical PC featuring Intel(R) UHD Graphics 620. The PC is equipped with an Intel(R) Core(TM) i5-8365U CPU processor clocked at 1.60 GHz, containing 8 CPUs, and backed by 16.0 GB of RAM. The operating system utilized is Windows 10 Enterprise 64-bit.

\subsubsection{Simulated data}
\begin{figure}
\centering
\includegraphics[width=3in, height=2.7in]{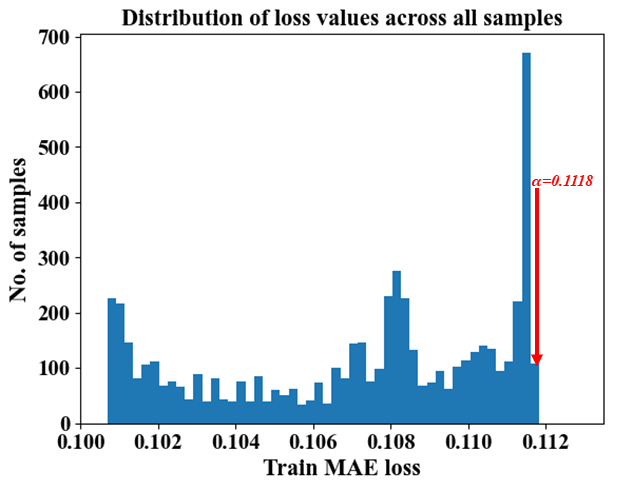}
  \caption{Distribution of reconstruction error of the model on train data. Each bin represents the number of segments of data having a given MAE loss. The highest reconstruction error is chosen as the threshold($\alpha$) for detecting the faults.}
  \label{fig:hist_train}
\end{figure}

\begin{figure*}[h!]
\centering
\captionsetup[subfigure]{font=small}
\subfloat[Phase A]{\includegraphics[width=0.3\textwidth, height=0.28\textwidth]{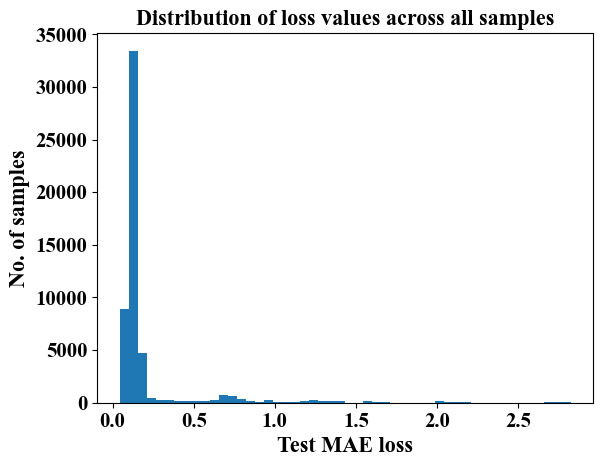}}\;
\subfloat[Phase B]{\includegraphics[width=0.3\textwidth, height=0.28\textwidth]{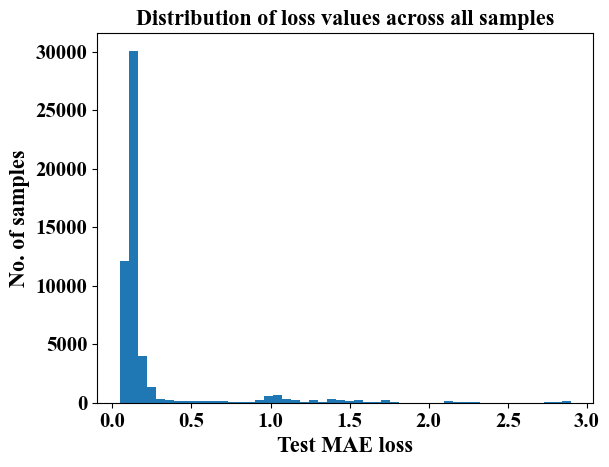}}\;
\subfloat[Phase C]{\includegraphics[width=0.3\textwidth, height=0.28\textwidth]{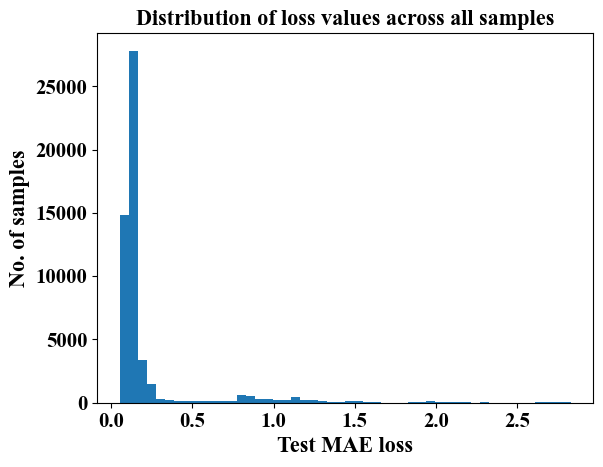}}\;
\caption{Distribution of reconstruction error of the model on test data of all three phases.}
\label{fig:hist_test}
\end{figure*}

\begin{figure*}[h!]
\centering
\captionsetup[subfigure]{font=small}
\subfloat[Phase A]{\includegraphics[width=0.3\textwidth, height=0.28\textwidth]{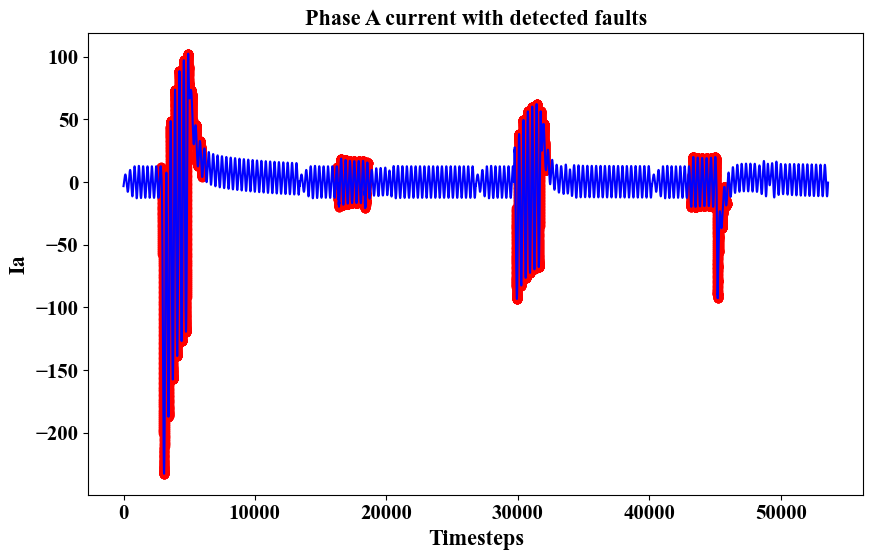}}\;
\subfloat[Phase B]{\includegraphics[width=0.3\textwidth, height=0.28\textwidth]{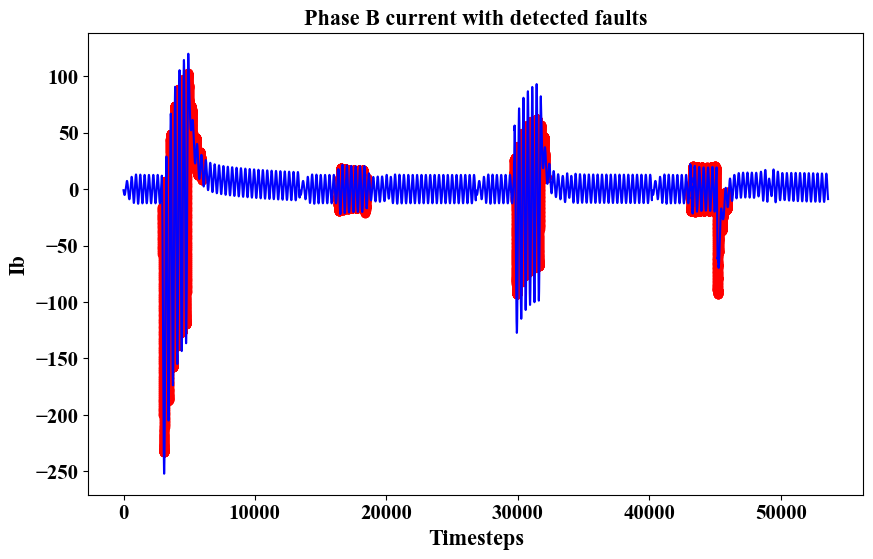}}\;
\subfloat[Phase C]{\includegraphics[width=0.3\textwidth, height=0.28\textwidth]{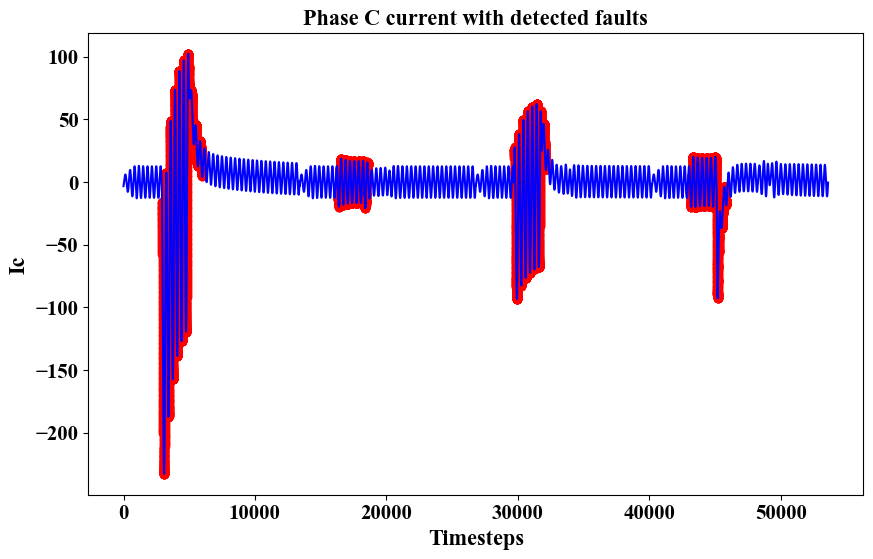}}\;
\caption{Current signals of three phases with highlighted fault segments detected by the model}
\label{fig:det_sim}
\end{figure*}

\begin{figure*}[h!]
\centering
\captionsetup[subfigure]{font=small}
\subfloat[Simulated data]{\includegraphics[width=0.48\textwidth]{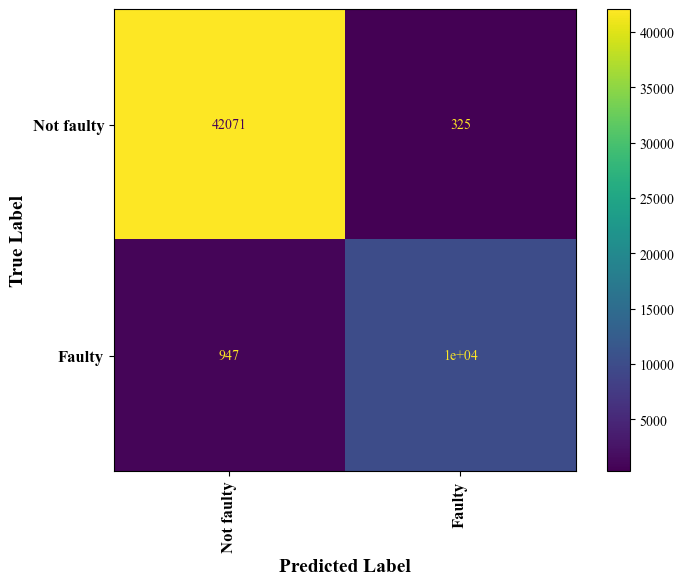} \label{fig:a}}\;
\subfloat[Public data]{\includegraphics[width=0.48\textwidth]{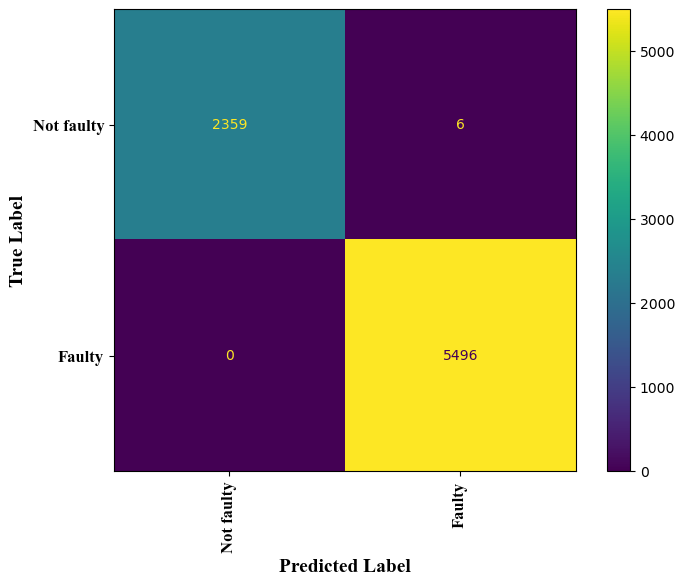}} \label{fig:b}\;
\caption{Confusion matrices}
\label{fig:cf_sim}
\end{figure*}

The autoencoder model was trained on the current signal of phase A under no-fault conditions. After the model was trained, it was given the phase A normal current signal for reconstruction. The reconstruction error of each sample in the signal was noted and the highest reconstruction error was taken as the threshold value $\alpha$. Fig. \ref{fig:hist_train}
shows the distribution of the reconstruction error across different samples when the model was given the training data for reconstruction. Next, the faulty current signal of each phase was given to the model for reconstruction separately. The segments for which the reconstruction error was greater than the threshold ($\alpha$) was detected as the fault segment. Fig. \ref{fig:hist_test}
shows the distribution of the reconstruction error across different samples when the model was given the test data, i.e., the signal with fault segments for reconstruction. Fig. \ref{fig:det_sim}
shows the current signals of all three phases with detected fault segments by the model. In this process, the model identified each data point as either faulty or not faulty, and the resulting confusion matrix is shown in Fig. \ref{fig:cf_sim}. Table \ref{tab:eval_sim} shows the evaluation metrics when the model was evaluated on a simulated dataset.

% \begin{figure}
% \centering
% \includegraphics[width=3in, height=2.7in]{Figures/cf_simulated.png}
%   \caption{Confusion Matrix on simulated data}
%   \label{fig:cf_sim}
% \end{figure}

\begin{table}[]
\centering
\caption{Evaluation metrics for simulated and public dataset}
\begin{tabular}{ccc}
\toprule
Metric  & Simulated data  & Public data\\
\midrule
Accuracy & 97.62 & 99.92     \\
Precision  &  96.85 & 99.89   \\
Recall &  91.35 & 1.00 \\
Specificity & 99.23 & 99.74 \\
F1-score &    94.02 & 99.94 \\
\bottomrule
\end{tabular}
\vspace{2mm}
\label{tab:eval_sim}
\end{table}

\subsubsection{Public data}
To further evaluate the model efficiency, we tested the model on a publicly available dataset in Kaggle, which consisted of around 7600 data points. Similar to the simulated dataset, the autoencoder model was trained on the current signal of phase A with no faults. The reconstruction error of each sample in the signal was noted, and the highest reconstruction error was taken as the threshold value. Then, the faulty current signal of each phase was given to the model for reconstruction separately. The segments for which the reconstruction error was greater than the threshold was detected as the fault segment. Fig. \ref{fig:hist_test} shows the distribution of the reconstruction error across different samples when the model was given the test data, i.e., the signal with fault segments for reconstruction. Fig. \ref{fig:det_sim}
shows the current signals of all three phases with detected fault segments by the model. The resulting confusion matrix is shown in Fig. \ref{fig:cf_sim}. Table \ref{tab:eval_sim} shows the evaluation metrics when the model was evaluated on the public dataset. Our model's performance was compared against the performance of other traditional ML models that were available in Kaggle and was found to be performing at par or even better than other ML models. Table \ref{tab:comp} shows the comparison scores. 

Thus by leveraging the ability of autoencoders to capture complex patterns in data and detect deviations from learned normal patterns, they can be effective tools for anomaly detection in time series data.

\begin{figure*}[h!]
\centering
\captionsetup[subfigure]{font=small}
\subfloat[Phase A]{\includegraphics[width=0.3\textwidth, height=0.28\textwidth]{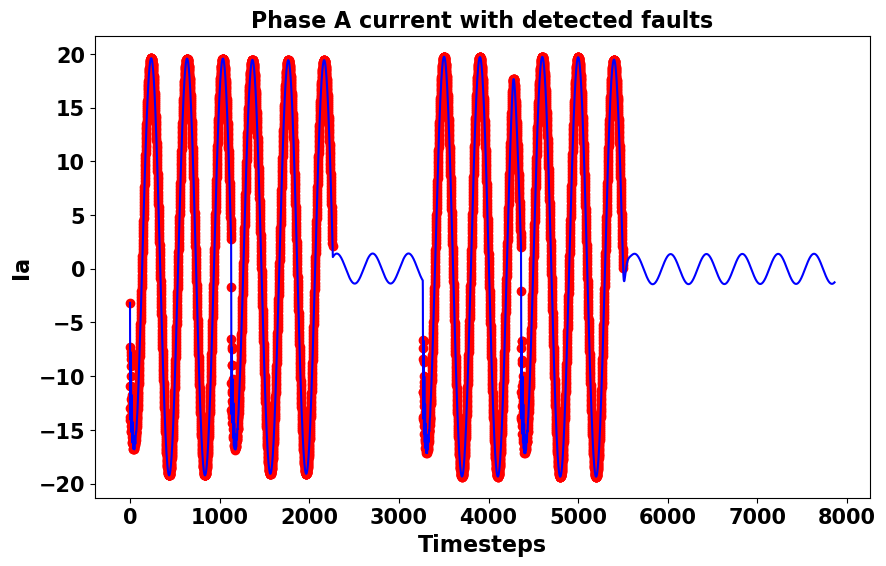}}\;
\subfloat[Phase B]{\includegraphics[width=0.3\textwidth, height=0.28\textwidth]{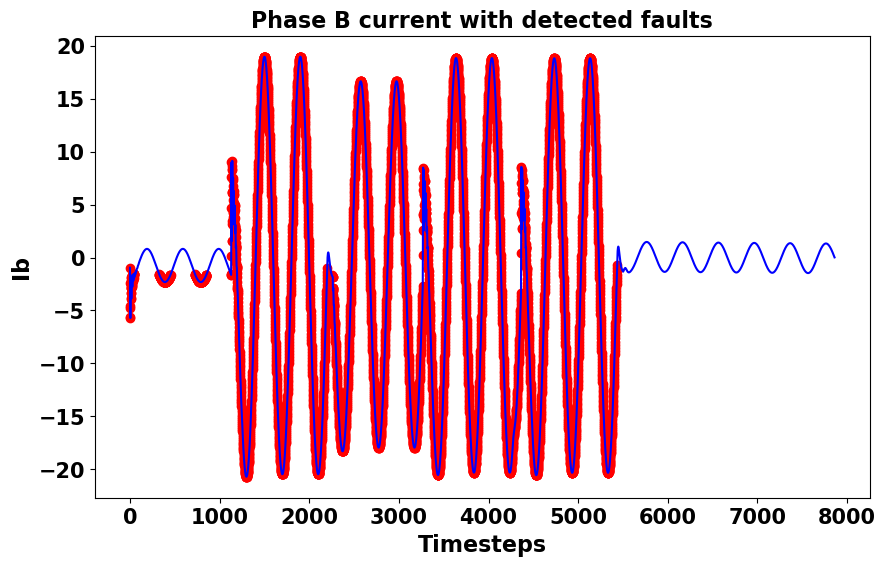}}\;
\subfloat[Phase C]{\includegraphics[width=0.3\textwidth, height=0.28\textwidth]{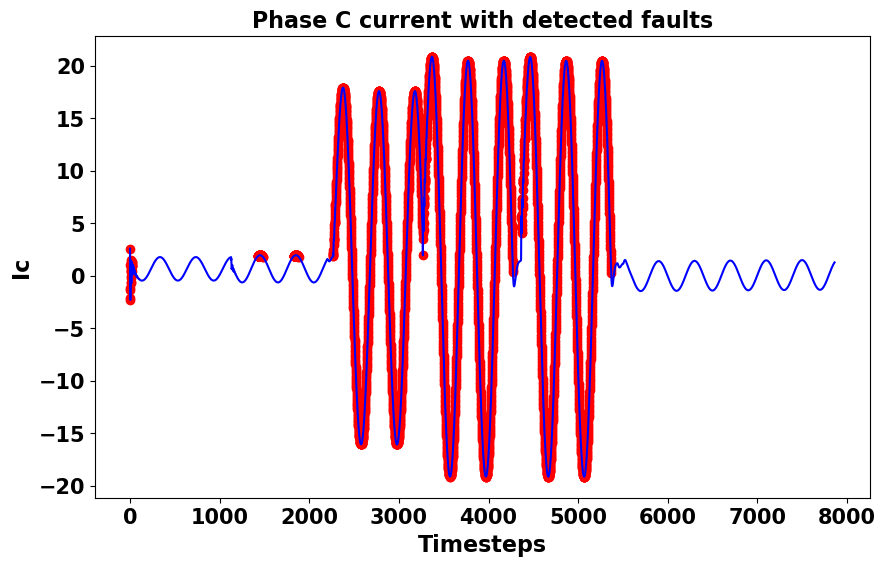}}\;
\caption{Current signals of three phases with highlighted fault segments detected by the model on public dataset}
\label{fig:det_pub}
\end{figure*}

\begin{table}[]
\centering
\caption{Comparision of our model performance with other ML models}
\begin{tabular}{lc}
\toprule
Models  & Accuracy  \\
\midrule
Logistic Regression \cite{ref_code} & 60.11     \\
Support Vector Machine \cite{ref_code} &  99.66  \\
K Neighbors Classifier \cite{ref_code} &  99.34 \\
Proposed Model & 99.92 \\
\bottomrule
\end{tabular}
\vspace{2mm}
\label{tab:comp}
\end{table}

\section{Conclusion}
In this work, we introduced a methodology that uses CNN based autoencoder model for efficient electrical fault detection. A large fault dataset was simulated with four different types of faults. The proposed method showed consistency in detecting each fault. For further evaluation, the model was tested on a publicly available dataset that contained various types of faults and was found to give a very high accuracy for detecting faulty points and performed better than other traditional ML models. Thus, the proposed methodology could offer practical benefits in terms of enhancing fault detection accuracy, reducing downtime, and improving system reliability. Future research could focus on testing the proposed methodology for a wide range of events and exploring its applicability in other domains beyond electrical systems. Additionally, integrating real-time data and adaptive learning algorithms could further enhance its capabilities.

\label{section:Conclusion}

\section*{Acknowledgments}
Sidharthenee Nayak and K. Victor Sam Moses Babu would like to thank ABB Ability Innovation Centre, Hyderabad, for their financial support in carrying out their research work. The author Mayukha Pal would like to thank the ABB Ability Innovation Center, Hyderabad, for their support in this work.

\section*{Data and Code Availability}
The public dataset for our analysis is from a free public database (https://www.kaggle.com/datasets/esathyaprakash/electrical-fault-detection-and-classification/data). The codes that support the findings of this study are available from the corresponding author upon reasonable request.


% \bibliography{sn-bibliography}% common bib file

\end{document}